\documentclass[twocolumn,secnumarabic,amssymb, nobibnotes, superscriptaddress, nofootinbib, aps, prd]{revtex4-2}
\usepackage{natbib}
\usepackage{xcolor}

\usepackage{hyperref}     
\usepackage{amsmath}
\usepackage{adjustbox}
\usepackage{graphicx}   
\usepackage{amsfonts}      
\usepackage{nicefrac}    
\usepackage{microtype}
\usepackage{cleveref}
\usepackage{colortbl}
\usepackage{makecell}
\usepackage{dsfont}
\usepackage{array}
\usepackage{orcidlink}

\begin{document}

\title{Comprehensive Analysis of Constructing Hybrid Stars with an RG-consistent NJL Model}

\author{Jan-Erik Christian~\orcidlink{0000-0003-0598-1621}}
\email{jan-erik.christian@uni-hamburg.de}
\affiliation{Hamburger Sternwarte, University of Hamburg, Gojenbergsweg 112, 21029 Hamburg, Germany}    
\author{Ishfaq Ahmad Rather~\orcidlink{0000-0001-5930-7179}}
\email{rather@astro.uni-frankfurt.de}
\affiliation{Institut f\"{u}r Theoretische Physik, Goethe Universit\"{a}t, 
	Max-von-Laue-Str.~1, D-60438 Frankfurt am Main, Germany}
\author{Hosein Gholami~\orcidlink{0009-0003-3194-926X}}
\email{mohammadhossein.gholami@tu-darmstadt.de}
\affiliation{Technische Universit\"{a}t Darmstadt, Fachbereich Physik, Institut f\"{u}r Kernphysik, Theoriezentrum, Schlossgartenstr.~2, D-64289 Darmstadt, Germany}
\author{Marco Hofmann~\orcidlink{0000-0002-4947-1693}}
\email{marco.hofmann@tu-darmstadt.de}
\affiliation{Technische Universit\"{a}t Darmstadt, Fachbereich Physik, Institut f\"{u}r Kernphysik, Theoriezentrum, Schlossgartenstr.~2, D-64289 Darmstadt, Germany}

\date{\today}

\begin{abstract}
    In this work, we investigate the properties of hadronic and quark matter that would allow for a first order phase transition between them within neutron stars.
To this end, we use a parameterizable Relativistic Mean-Field (RMF) description for the hadronic phase and a Renormalization Group-consistent Nambu-Jona-Lasino (RG-NJL) model for the quark phase. This also enables us to consider sequential phase transitions involving a two-flavor color-superconducting (2SC) and a color-flavor-locked (CFL) phase. We find large ranges for all parameters that facilitate a phase transition, even when constrained by current astrophysical data. 
We further attempt to filter out stars with a high chance of detectability by mass-radius measurement, i.e., stars with identical mass but different radii, so-called twin stars. However, we find that such configurations are outside the constrained parameter spaces. Instead, most of the mass-radius relations that feature a phase transition appear to be indistinguishable from a purely hadronic description.
\end{abstract}

\maketitle

\section{Introduction}
Neutron stars are among the densest objects in the known universe. As such, they can provide unique insights into the behavior and composition of bulk matter under extreme conditions. The equation of state (EoS), which describes such matter by connecting energy density and pressure, is strongly linked to the mass, radius, and tidal deformability of neutron stars and can therefore be probed by measuring those quantities.
As such, astrophysical observations play a significant role in constraining the EoS of neutron star matter \cite{Annala:2017llu, Landry:2020vaw, Dietrich:2020efo, Chatziioannou:2020pqz, Annala:2021gom}. The first significant reduction of possible EoSs in recent history was initiated by the detection of several neutron stars with masses around $2\,M_\odot$ \cite{Demorest:2010bx, Antoniadis:2013pzd, Fonseca:2016tux, Cromartie:2019kug, Romani:2022jhd}. This high mass necessitates that the pressure rises steeply as a function of energy density. In other words, the EoS is required to be stiff. While neutron star masses can be accurately measured, determining their radii remains more challenging. However, recent observations from the NICER collaboration have provided valuable constraints on neutron star radii \cite{Miller:2019cac, Riley:2019yda, Raaijmakers:2019qny, Miller:2021qha, Riley:2021pdl, Raaijmakers:2021uju, Salmi:2024aum, Dittmann:2024mbo, Choudhury:2024xbk}, and can be used to further constrain the EoS. 
However, the most influential data point comes from the detection of GW170817, the first gravitational wave signal from a binary neutron star merger with an electromagnetic counterpart \cite{TheLIGOScientific:2017qsa, Abbott:2018exr, Abbott:2018wiz, Abbott_2020}. This event puts stringent limits on the tidal deformability of a $1.4\,M_\odot$ neutron star, thereby constraining the EoS. Notably, the tidal deformability points to a softer EoS, in contrast to the $2\,M_\odot$ constraint.\\

While a significant amount of EoSs in the literature contain only hadronic matter, it has been shown that a phase transition to quark matter can help ease this apparent tension between the maximum mass and tidal deformability constraint \cite{Paschalidis:2017qmb, Christian:2018jyd, Montana:2018bkb, Sieniawska:2018zzj, Christian:2019qer}. Such an EoS could lead to a neutron star with a quark matter core and a hadronic envelope, a so-called hybrid star \cite{Ivanenko:1965dg,Itoh:1970uw,Alford:2004pf,Coelho:2010fv,Chen:2011my,Masuda:2012kf,Yasutake:2014oxa,Zacchi:2015oma,Xie:2020rwg}. Hybrid stars cannot inherently be distinguished from purely hadronic neutron stars \cite{Alford:2004pf}; however, if the phase transition is suitably strong, certain indicators would be detectable in mass, radius, and tidal deformability. 
One of the more extreme versions of such an indicator would be a so-called twin star configuration 
\cite{Kampfer:1981yr,Glendenning:1998ag,Schertler:2000xq,SchaffnerBielich:2002ki,Zdunik:2012dj,Blaschke:2015uva,Zacchi:2016tjw,Blaschke:2019tbh,Tan:2020ics}. 
Twin stars require the existence of two stable branches in the mass-radius diagram, separated by an unstable region. They are defined as two stars with the same mass but different radii. This configuration cannot be achieved without a phase transition. Apart from the strength of the phase transition, the choice of the quark matter model plays a significant role in the construction of hybrid and twin stars. 
Specifically, a stiff quark matter model is more likely to exhibit a disconnected second branch than a soft one \cite{Alford:2013aca}. Therefore, modeling the quark phase with a constant speed of sound (CSS) approach with high values for the sound speed is 
a popular method \cite{Alford:2013aca,Alford:2015dpa,Alford:2017qgh,Christian:2017jni,Christian:2019qer,Christian:2023hez,Li:2024lmd}. Here we use a quark matter model with a stronger basis in microphysics.\\

We employ a relativistic mean-field (RMF) EoS \cite{PhysRev.98.783, Duerr56,Walecka74, Boguta:1977xi, Serot:1984ey, Mueller:1996pm, Typel:2009sy, ToddRutel:2005fa, Chen:2014sca, Fattoyev:2017jql,Hornick:2018kfi} to describe hadronic matter. This is motivated by its high degree of parameterizability combined with its microphysically well-motivated approach. We combine it with a color-superconducting (CSC) quark matter description. This type of quark matter is characterized by Cooper pairing between the quarks, leading to a nonzero diquark energy gap in the quasiparticle excitation spectrum \cite{Barrois:1977xd,bailin1979superfluid,Bailin:1983bm,Alford:1997zt,Rapp:1997zu,Alford:1998mk,Son:1998uk, Schafer:1999jg}. In the two-flavor color-superconducting (2SC) phase, up-and-down quark flavors of two colors (e.g., red and green) participate in the pairing, while one color (blue) and the strange quarks (if abundant) remain unpaired \cite{Alford:1999pa}. In the color-flavor locked (CFL) phase, quarks of all colors and flavors participate in the pairing \cite{Alford:1998mk, Schafer:1999fe, Shovkovy04}. The Nambu-Jona-Lasinio (NJL) model is a microscopic model for quark matter that allows for a self-consistent calculation of chiral symmetry breaking and sequential CSC phases, including the 2SC and CFL phases. It has been widely used and extended to describe quark matter in compact star physics, e.g. \cite{Baldo:2002ju, Buballa:2003et, Klahn:2006iw,Pagliara:2007ph, Bonanno:2011ch, Klahn:2013kga, Baym:2017whm,Ranea-Sandoval:2017ort, Alaverdyan:2020xnv, Alaverdyan:2022foz, PhysRevC.110.045802, Yuan:2025dft}.
For the present work, it is important that quark matter at the core densities of the most massive star configurations ($n_B$ up to $\sim10$ times nuclear saturation density) can be reliably calculated. To ensure this, we use the three-flavor Renormalization Group-consistent NJL (RG-NJL) model \cite{Gholami:2024ety, Gholami:2024diy} to ensure that our calculations at high densities are not spoiled by cutoff artifacts. We assume the transition between hadronic and quark matter to be of first order and model it with a Maxwell construction.\\

The role of CSC in hybrid stars has become an active field of research in recent years, with various quark matter descriptions applied, such as non-local NJL models \cite{Shahrbaf:2021cjz, Blaschke:2022egm, Ivanytskyi:2024zip}, confining relativistic density functional approaches~\cite{Ivanytskyi:2022oxv, Ivanytskyi:2022bjc, Gartlein:2023vif, Gartlein:2024cbj}, and parameterized models incorporating strong perturbative QCD corrections~\cite{Zhang:2020jmb}.\\ 

This work aims to provide a comprehensive analysis of the possible parameter space for vector and diquark couplings within an RG-consistent framework.
We constrain the parameter space of the RMF-NJL hybrid model with measurements from NICER, the tidal deformability measurement of GW170817, and by requiring that the EoS allows for neutron stars with a maximum mass of at least $2\,M_\odot$. We find that a phase transition to quark matter prefers a stiff hadronic EoS. This means that the tidal deformability constraint imposes significant limitations on possible parameter variations for both phases. Nevertheless, a large amount of hybrid EoSs are within astrophysical constraints. We also consider possible twin star configurations, showing that the $2\,M_\odot$ constraint in combination with the tidal deformability constraint from GW170817 reduces the parameter space for twin stars to such an extent that they can effectively be ruled out, within our approach.\\ 

Our work is organized as follows: In Section (\ref{Sec:EoS}) we provide a detailed description of the equation of state for the hadron and quark phase, as well as the hadron-quark transition. In Section (\ref{Sec:Results}) we present the results from an extensive analysis of $\sim\,$700,000 parameter combinations, incorporating the mass constraint, as well as NICER and GW170817 observations. Finally, in Section (\ref{Sec:disc}) we discuss our findings.\\

\section{Equation of state}\label{Sec:EoS}

\subsection{Hadronic Phase}
For the crust EoS, we employ the well-known Baym-Pethick-Sutherland (BPS) EoS \cite{Baym:1971pw, Negele:1971vb}. The hadronic part of the core is described using a relativistic mean-field (RMF) parametrization \cite{PhysRev.98.783, Duerr56,Walecka74, Boguta:1977xi, Serot:1984ey, Mueller:1996pm, Typel:2009sy, ToddRutel:2005fa, Chen:2014sca} within the limits explored in \cite{Hornick:2018kfi}. We refer to this as the Frankfurt-Barcelona (FB) EoS. In this setup, the symmetry energy $J$, the slope parameter $L$, and the effective mass $m^*/m$ are fixed at saturation density $n_0$ in a way that is compatible with chiral effective field theory calculations. Furthermore, we vary these parameters within the ranges allowed by experiments \cite{Lattimer:2023rpe}, allowing us to provide a full picture of the possible parameter space for phase transitions to our quark model. Small effective masses correspond to stiff EoSs \cite{Boguta:1981px}, where stiff refers to the pressure as a function of energy density rising quickly when compared with a soft EoS. Stiff EoSs have higher masses, radii, and tidal deformabilities than soft EoSs. As a result, our hadronic model requires a value of $m^{*}/m > 0.65$ to be consistent with the tidal deformability of GW170817 \cite{Abbott:2018wiz}. A value of $m^{*}/m = 0.65$ is only within the limits if the other parameters, fixed at saturation density, are adjusted such that they contribute to a soft EoS \cite{Hornick:2018kfi, Ghosh:2021bvw}. 
However, following the detection of GW170817, it has been shown that its constraining power on the hadronic EoS can be mitigated with a suitable phase transition to quark matter \cite{Paschalidis:2017qmb, Christian:2018jyd, Montana:2018bkb, Sieniawska:2018zzj, Christian:2019qer}. This is the case because hybrid stars are more compact than purely hadronic neutron stars, allowing previously ruled-out hadronic phases to regain compatibility. With this in mind, we vary the effective mass within the entire range $m^{*}/m = 0.55-075$ laid out by \cite{Hornick:2018kfi} instead of limiting us to the range constrained by GW170817. The slope parameter is varied between $L=40-60\,\mathrm{MeV}$, the symmetry energy between $J = 30-32\,\mathrm{MeV}$, and finally we vary the value of the nuclear saturation density $n_0 = 0.15-0.16\,\mathrm{fm^{-3}}$. The influence of $J$, $L$, and $n_0$ on mass, radius, and tidal deformability is minimal but not negligible. In Section \ref{Sec:Results},
we show that the ability to combine a given quark EoS with the hadronic envelope is affected by these parameters as well.

\subsection{Quark Phase}
\label{SubSec:Quarkmodel}

For the quark phase, we employ the three-flavor Renormalization Group-consistent Nambu-Jona-Lasinio (RG-NJL) model with a diquark interaction, allowing for the formation of homogeneous color-superconducting condensates. The NJL model is non-renormalizable, implying that the model's regularization dependence cannot be absorbed into renormalized quantities. In the literature, the NJL model is typically regularized with a sharp three-momentum cutoff $\Lambda'$ which is fitted to vacuum properties. The RG-consistent description of the NJL model \cite{Gholami:2024diy} removes artifacts of the conventional regularization and allows for a consistent investigation of the phase structure at high chemical potentials \cite{Braun:2018svj}. 

 As discussed in Ref.~\cite{Gholami:2024diy}, using a conventional cutoff regularization can introduce unphysical artifacts. Even evaluating the model at chemical potentials as small as 50\% of the cutoff $\Lambda'$ can lead to this effect. To prevent this issue, a renormalization-group consistent regularization scheme was proposed in Ref.~\cite{Braun:2018svj} and adapted for a neutral three-flavor NJL model with color superconductivity in Ref.~\cite{Gholami:2024diy}. The concept of RG-consistency can be straightforwardly formulated in the context of the functional renormalization group, in which the full quantum effective action is calculated by solving a flow equation for a scale-dependent average effective action, which is initialized at some initial ultraviolet scale $\Lambda$ \cite{Wetterich:1992yh}. RG-consistency then requires that the initial scale $\Lambda$ for the flow equation is set sufficiently high such that the full quantum effective action becomes independent of $\Lambda$ in the infrared limit \cite{Braun:2018svj}. In the context of the mean-field approximation, this large initial scale effectively acts as a high-momentum cutoff. For further details on the importance of RG consistency in mean-field models, see \cite{Braun:2018svj} and \cite{Gholami:2024diy}. 
In short, the RG-consistent regularization employs a large cutoff scale $\Lambda$ while the divergent vacuum contributions are regularized with the traditional cutoff $\Lambda'$. RG consistency is obtained for $\Lambda \gg \Lambda'$.
Additionally, when diquark condensates are present, new medium-dependent divergences emerge, which must be subtracted by appropriate counterterms to ensure that the physical results remain independent of the large cutoff scale. These counterterms can be motivated by different renormalization schemes.
In this work, we employ the so-called \textit{massless scheme,} which was also used for calculating a phase diagram and the EoS in Ref.~\cite{Gholami:2024diy} and Ref. \cite{Gholami:2024ety}.\\

The different contributions to the Lagrangian density along with the effective potential in the RG-consistent regularization for the model are outlined in detail in Ref.~\cite{Gholami:2024ety}. The grand potential per volume $\Omega (\mu_B, T)$ is a function of the baryon chemical potential $\mu_B$ and temperature $T$. The pressure $P$ is obtained as  
\begin{eqnarray}
    P&=&-(\Omega-\Omega_0)-B,\label{pressnjl}
\end{eqnarray}
wit $\Omega_0=\Omega(\mu_B=0,T=0)$.
The parameter $B$ (in units of MeV$\cdot$fm$^{-3}$) acts as an additional bag constant, which fixes the pressure in vacuum to $P(\mu_B=0, T=0)=-B$. For thermodynamic consistency, the energy density $\epsilon$ is increased for nonzero $B$ as $\epsilon\to \epsilon+B$.\\

Overall, the quark matter model includes three free parameters, which are varied to determine the EoS. We then assess its compatibility with the astrophysical data, ultimately constraining them. These are the diquark coupling $\eta_D$, the vector coupling $\eta_V$, and the additional bag constant $B$. The stiffness and the composition of the quark phase at a fixed density can be varied by a variation of the couplings $\eta_D$ and $\eta_V$. In an earlier study, we investigated the stability of hybrid stars with a 2SC-CFL phase transition \cite{Gholami:2024ety}. For this study, we employed parameter ranges of $\eta_V = 0.00 - 1.50$, and $\eta_D = 0.90 - 1.80$. Nearly all of the maximum-mass configurations of the parameter sets surpassing $2\,M_\odot$ contained a CFL quark phase at their center. Note, however, that these calculations were done exclusively with a vanishing bag constant $B = 0\,\mathrm{MeV/fm^3}$.\\

The NJL model lacks confinement and is only used for the high-density quark matter description of our hybrid model.
Consequently, the vacuum pressure calculated within the NJL model does not necessarily have to match the vacuum pressure derived from the RMF model. In this work, we use the freedom to choose a nonzero $B$ to effectively change the density of the hadron-quark phase transition. This is a standard procedure in the literature \cite{Rather:2020lsg, Burgio:2002sn, Steiner:2000bi, Rather:2020mlz}. Increasing $B$ lowers the pressure of quark matter at every point and moves the hadron-quark phase transition to higher densities.\\

Apart from $\eta_V$, $\eta_D$, and $B$, the remaining coupling constants of the model and the bare quark masses are fitted to the pseudoscalar meson octet in vacuum \cite{Rehberg:1995kh}.
They take the same values as in our previous work \cite{Gholami:2024diy}.
Since the present model is RG-consistent, it can be evaluated at arbitrary high scales with the condition that the large cutoff scale $\Lambda$ is chosen sufficiently large. For this work, $\Lambda=10 \Lambda'=6023$\,MeV is found to be sufficiently high to study the densities reached in neutron stars.\\

\subsection{Phase Transition}
The hadron-quark phase transition in this work is modeled using a Maxwell construction. In this setup, the baryon chemical potential and pressure are kept constant at the point of transition $p_{trans}$, but the (energy) density is expected to jump. A prerequisite for this construction is a ``strong'' phase transition, which is more likely to lead to noticeable changes in the mass-radius relation than the more involved Gibbs construction, where the charge chemical potential is conserved across the mixed phase as well \cite{Hempel:2009vp}. As the Maxwell construction requires a point where $\mu_B(p)_{HM} = \mu_B(p)_{QM}$, a match between a given hadronic EoS and a quark EoS is not always possible. Additionally, we imposed that a phase transition can only occur at densities larger than $n_0$ and smaller than $10\,n_0$. The upper-density limit is not physically motivated but is chosen to keep our phase transitions close to neutron star densities. For our description of hadronic and quark matter, the parameter choices affect a possible match to different degrees. The nuclear parameters $m^{*}/m$ and $L$, have the largest impact on the allowed values of $\eta_V$ and $\eta_D$. We discuss this in detail in Section~\ref{Sec:Results}.\\

The presence of a phase transition can lead to a significant impact on the mass-radius relation and therefore on astrophysical observables. This usually takes the form of a kink in the mass-radius relation or even a scenario where the sequence becomes unstable when $p_{central}=p_{trans}$ and regains stability at higher central pressures, forming a stable second branch containing hybrid stars. If such a second branch forms, two neutron stars with the same mass but different radii and tidal deformability can occur, i.e. twin stars. 
The significance of these twin stars lies solely in them being indicators for a phase transition that can be checked with mass, radius (and tidal deformability $\Lambda$) measurements. In the following, we will explore what conditions allow for a phase transition and which of these phase transitions lead to twin star configurations.\\

\section{Results}\label{Sec:Results}
\subsection{Matching Hadronic and Quark Phase}\label{Subsec:Matches}

In this work, we investigated over 700,000 parameter combinations, of which about 125,000 feature a possible transition point and 22,000 comply with astrophysical constraints. The ranges and steps in which our parameters were varied are listed in Table~\ref{Tab:Ranges}. We found that the last stable star for every parameter set either has quark matter in the CFL phase at its center or a purely hadronic core; there are no cases where the densest star features a 2SC center. However, for some EoSs there are less dense stars that can feature a pure 2SC quark core.\\

\begin{figure*}
		\centering				
		\includegraphics[width=0.8\textwidth]{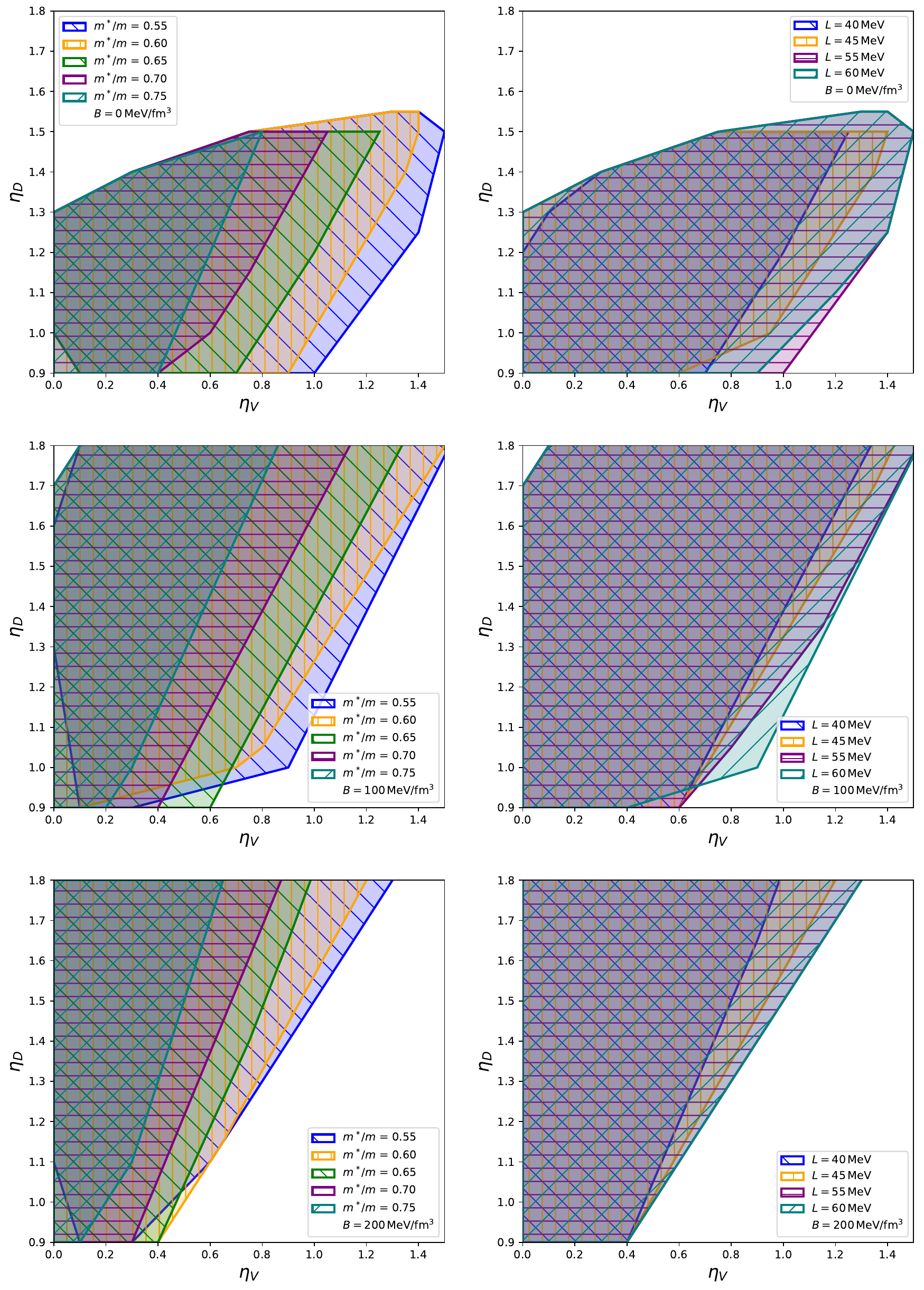}
		\caption{Left: The effective mass $m^{*}/m$ is kept constant, $n_0$, $J$, and $L$ are not fixed. The shaded areas represent the region where a hadronic EoS with the given $m^{*}/m$ can be matched with the corresponding quark matter parameters $\eta_V$ and $\eta_D$.
        Right: The slope parameter $L$ of the hadronic phase is kept constant, $n_0$, $J$, and $m^{*}/m$ are not fixed. The results of the first row are for a bag constant $B=0\,\mathrm{MeV/fm^3}$, which is increased in each descending row by $100\,\mathrm{MeV/fm^3}$.}
		\label{Fig:CAfM}
\end{figure*}

\begin{figure*}
		\begin{minipage}[t]{0.47\textwidth}		 		
  \includegraphics[width=\textwidth]{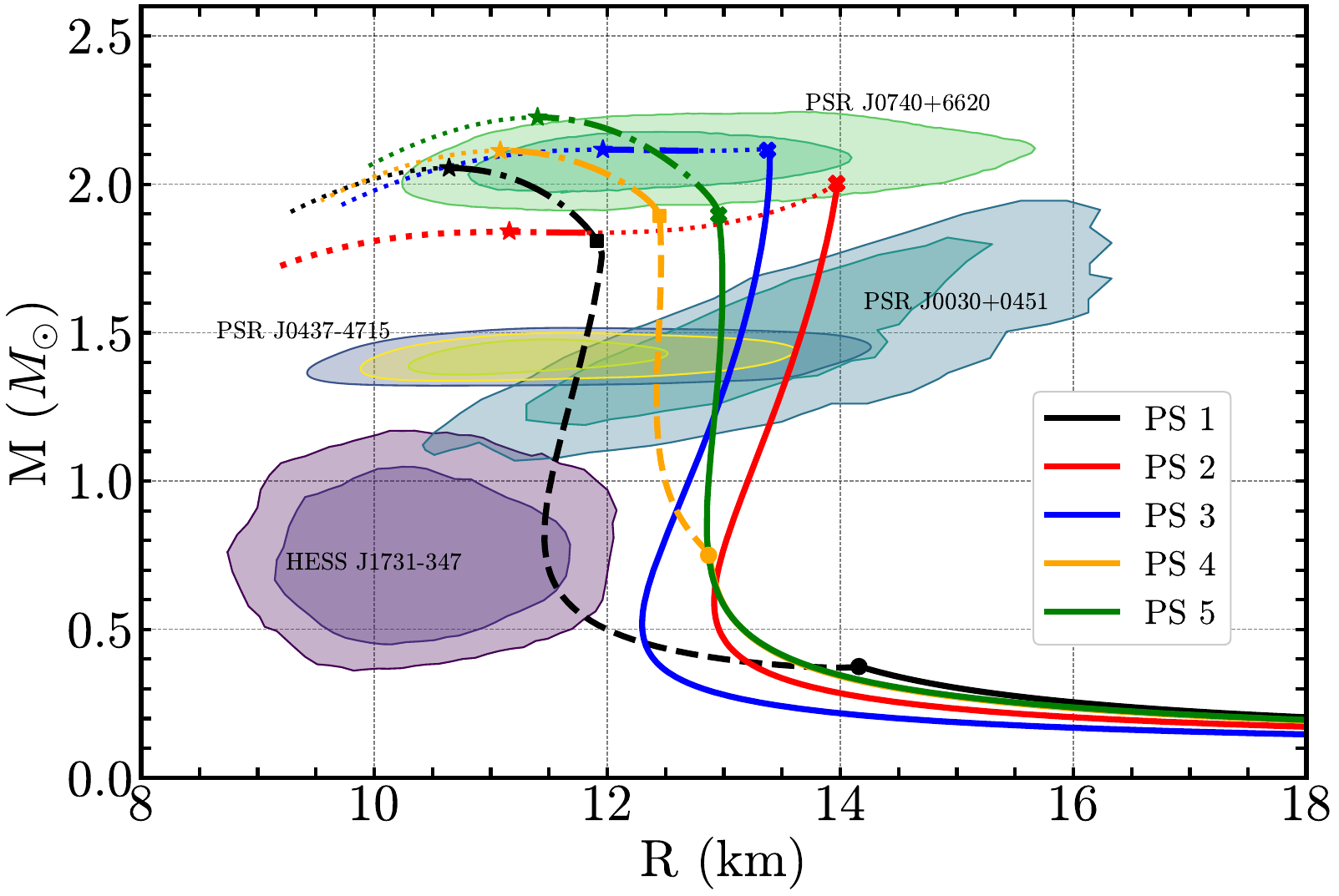}
			 	\end{minipage}
		 		\begin{minipage}[t]{0.47\textwidth}
			 		\includegraphics[width=\textwidth]{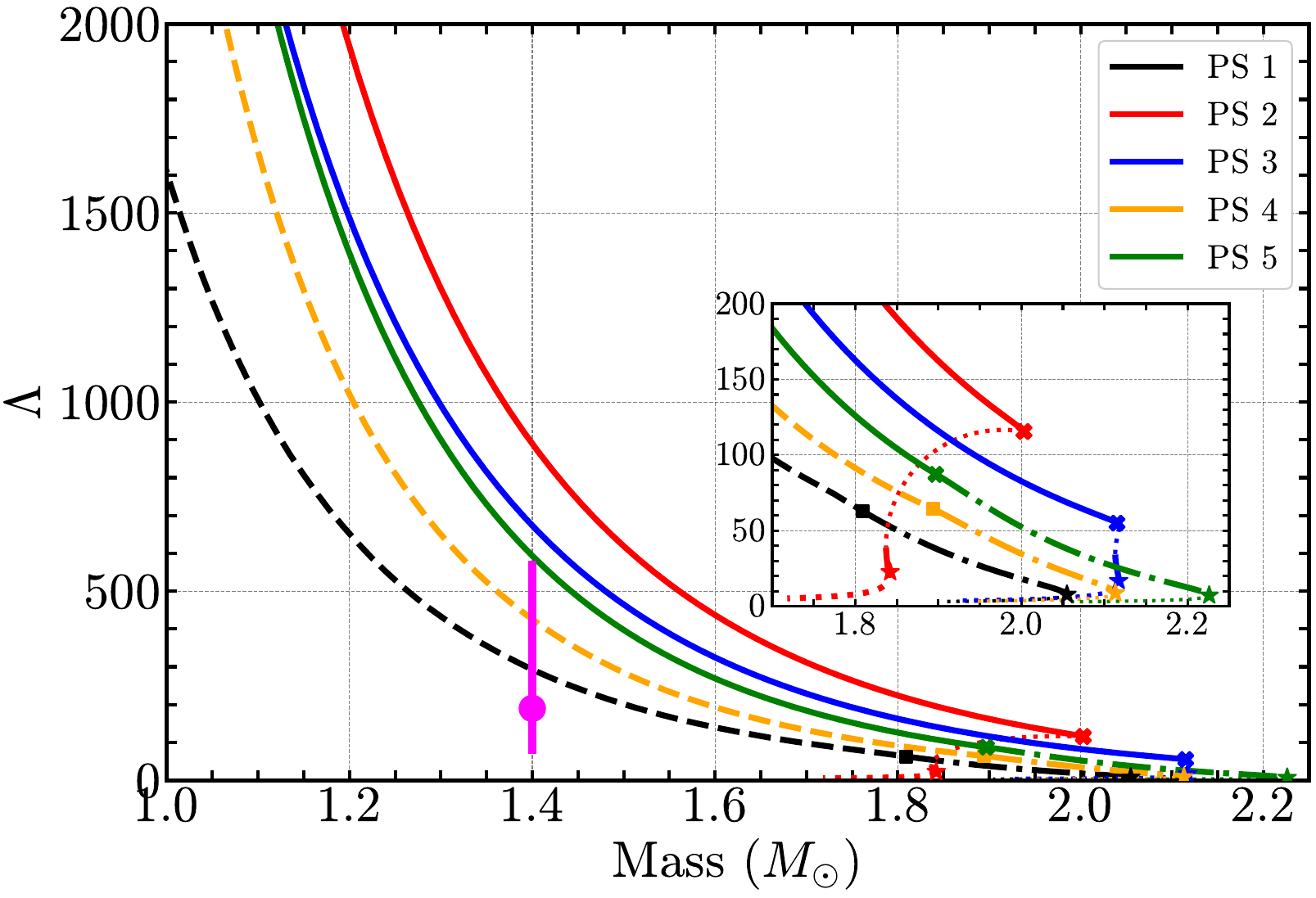}
			 	\end{minipage}
			 			\caption{Left: Mass-radius relation for hybrid stars with different hadronic and quark phase parameters as displayed in Table~\ref{Tab:MR_Parametersets}.
        Points on solid lines are stars with hadronic matter in the core, while points on the dashed and dot-dashed lines contain 2SC and CFL quark matter in the center, respectively. The star symbol indicates the densest stable star.
        The solid circle and square symbols mark the onset of the 2SC and CFL phases in the core. The cross symbol corresponds to a direct phase transition from hadronic matter to the CFL phase. The dotted line corresponds to the unstable region. The various shaded areas are $1\,\sigma$ and $2\,\sigma$
       credibility regions for mass and radius inferred from the analysis of the pulsar stated in the label.  
       Right: Corresponding dimensionless tidal deformability ($\Lambda$) - mass relation. 
       The solid magenta line represents the 90\% credibility constraint on the dimensionless tidal deformability at 1.4\,$M_{{\odot}}$ from the GW170817 measurement \cite{Abbott:2018exr}. The high-mass region is shown in more detail.}
		\label{Fig:MRL}	 	
     \end{figure*}

\begin{table}[h]
    \centering
    \caption{Parameter ranges and step sizes. The effective mass $m^*/m$, vector coupling $\eta_V$, and diquark coupling $\eta_D$ are dimensionless. Symmetry energy $J$ and slope parameter $L$ are in MeV. Saturation density $n_0$ and bag constant $B$ are in units of $\mathrm{fm^{-3}}$ and $\mathrm{MeV/fm^3}$, respectively. Step sizes for $\eta_V$ and $\eta_D$ vary for efficiency. $B$ uses steps of 10 $\mathrm{MeV/fm^3}$ up to 50 $\mathrm{MeV/fm^3}$, then steps of 50 $\mathrm{MeV/fm^3}$.}
    \label{Tab:Ranges}
    \begin{tabular}{l c c c}
        \toprule
        Parameter & Lower bound & Upper bound & Step Size \\
        \hline
        $m^*/m$ & 0.55 & 0.75 & 0.05 \\
        $L$ [MeV] & 40 & 60 & 5 \\
        $J$ [MeV] & 30 & 32 & 1 \\
        $n_0$ [$\mathrm{fm^{-3}}$] & 0.15 & 0.16 & 0.01 \\
        $\eta_V$ & 0.0 & 1.5 & var \\
        $\eta_D$ & 0.9 & 2.0 & var \\
        $B$ [$\mathrm{MeV/fm^3}$] & 0 & 200 & 10 - 50 \\
        \hline
    \end{tabular}
\end{table} 

The effects of the nuclear parameters $n_0$, $J$, $L$, and $m^*/m$ on the $\eta_V-\eta_D$ parameter space are shown in Fig.~\ref{Fig:CAfM} and Fig.~\ref{Fig:CAfC}. In these figures, areas are highlighted by color for each of which we fix one of the four nuclear parameters. 
For each point in such an area, at least one matching parameter set exists for fixed $\eta_V$ and $\eta_D$ as well as the fixed nuclear parameter. For example, at least one hadronic parameter set with an effective mass of $m^*/m=0.65$ can be combined with a quark phase that is characterized by $\eta_V=0.80$ and $\eta_D = 1.30$. However, we cannot assume that all hadronic sets with $m^*/m=0.65$ can be combined with such a quark phase.\\ 

Before we begin with the parameter variations in Fig.~\ref{Fig:CAfM}, we note that we found the choice of symmetry energy $J$ to be inconsequential for the possibility of matching a hadronic EoS with a quark matter parameter set. The value of saturation density $n_0$ is only slightly more relevant. By increasing $n_0 = 0.15\,\mathrm{fm^3}$ to $n_0 = 0.16\,\mathrm{fm^3}$ we noticed a decrease in the allowed $\eta_V$ range of 0.1 at most, while the range of $\eta_D$ is unaffected.\\

The parameters $m^*/m$ and $L$ have a much larger impact on the possibility to match hadronic and quark matter than $n_0$ and $J$. This becomes clear in Fig.~\ref{Fig:CAfM}, where we show the $\eta_V-\eta_D$ space that has at least one parameter set with a given effective mass $m^*/m$ (left) or a given slope parameter $L$ (right) available to them. In addition, Fig.~\ref{Fig:CAfM} has three rows, where $B$ is increased in each row by $100\,\mathrm{MeV}$, starting at $B=0\,\mathrm{MeV}$ in the first row. This allows us to visualize not only the effect of $m^*/m$ and $L$, but $B$ as well.\\

We find that the effective mass and slope parameters affect the $\eta_V-\eta_D$ space in opposite ways. An increase in $m^*/m$ reduces the parameter space while an increase in $L$ enlarges it. What is common to these variations is a stiffening of the hadronic EoS. This is also the case for decreasing $n_0$. It is therefore clear that stiff EoSs are more easily combined with a quark phase than soft ones, as one would expect.
We further note that the $\eta_V$-range is strongly affected by softening the hadronic EoS, where much smaller values are required to make a match possible. The hadronic EoSs influence on the allowed range for $\eta_D$ is minimal. Instead, $\eta_D$ seems to be limited by the bag constant $B$, which in turn does not influence the allowed range for $\eta_V$. This is because increasing $B$ increases the transition density from hadronic to quark phase, which introduces parameter sets with $n_{trans}$ previously outside the investigated range.\\

In Fig.~\ref{Fig:MRL} we show the mass-radius, and the dimensionless tidal deformability - mass relations for a small number of representative EoSs. Since it is not possible to perfectly represent the entire parameter space we investigate, we chose to include examples to highlight the different possible types of phase transition, as well as EoSs that feature potentially detectable peculiarities, like twin stars. This also includes parameter sets (PS) that are not compatible with astrophysical observations, such as PS2 and PS3. The hadronic and quark phase parameters used to generate these mass-radius relations are listed in Table~\ref{Tab:MR_Parametersets}. The solid lines in Fig.~\ref{Fig:MRL} represent the purely hadronic stars. Stars on the dashed line feature a 2SC core and stars on the dot-dashed line have quark matter in the CFL phase at their center.
The transition point from hadronic to quark phase is marked by a solid dot. A square symbol marks the transition from the 2SC to the CFL phase. The cross symbol denotes a direct transition from hadronic matter to the CFL phase and the star symbol indicates the end of the mass-radius relation. The dotted line represents the unstable region.
On the left side of the figure, the mass-radius relation is shown in conjunction with the NICER constraints for PSR J0030+0451 \cite{Miller:2019cac, Riley:2019yda, Raaijmakers:2019qny}, PSR J0740+6620 \cite{Miller:2021qha, Riley:2021pdl, Raaijmakers:2021uju} and PSR J0437-4715 \cite{Choudhury:2024xbk,Salmi:2024aum, Dittmann:2024mbo} as well as the HESS J1731-347 data point \cite{Doroshenko2022}. PS1 has a very soft hadronic phase and an early onset of the quark phase at about $1.3\,n_0$. It is therefore a very extreme example. In addition, the mass-radius relation does not exhibit any clear signs of a phase transition taking place, it would be indistinguishable from a purely hadronic EoS. The only hint for its more exotic makeup is its placement at lower radii. Conversely, PS2 and PS3 lead to twin stars. In the case of PS2, the transition from the hadronic phase to the 2SC phase is accompanied by a sizable jump in energy density (latent heat) of $\Delta\epsilon = 370\,\mathrm{MeV/fm^3}$. This leads to the location of the hybrid star branch being moved to smaller masses and radii compared to PS3, which has a much smaller latent heat. For PS3, the hadronic phase transitions directly into the CFL phase with no 2SC phase present. The corresponding latent heat is small, placing the second branch at nearly the same mass as the hadronic maximum. Nevertheless, twin stars are present, which means that an observer unfamiliar with the EoS could infer the presence of a phase transition from the mass-radius relation.
The PS4 case also includes a kink in the mass-radius relation, but it is much less pronounced and is not resolvable with current observations. Like PS2, it features a phase transition from the hadronic to the 2SC phase, as well as a transition from the 2SC phase to the CFL phase. Finally, the mass-radius relation of PS5 is completely indistinguishable from a purely hadronic EoS, even though its early phase transition would result in exclusively hybrid stars being realized in nature.
The right side of Fig.~\ref{Fig:MRL} shows the tidal deformability as a function of mass for PS1-PS5. The solid magenta line represents the 90\% credibility constraint from the GW170817 measurement as $\Lambda_{1.4\,M_{\odot}}$ = 190$^{+390}_{-120}$ \cite{Abbott:2018exr}. Only PS1, PS4, and PS5 are within the constraint, which means that the two cases including twin stars are not consistent with GW170817.\\

\subsection{Applying Astrophysical 
Constraints}\label{Subsec:AstroConstraints}

In the following, we apply constraints from astrophysical data to all the available matches. Specifically, the masses and radii of PSR J0030+0451 \cite{Miller:2019cac,Riley:2019yda,Raaijmakers:2019qny}, PSR J0740+6620 \cite{Miller:2021qha,Riley:2021pdl,Raaijmakers:2021uju, Salmi:2024aum, Dittmann:2024mbo}, and PSR J0437-4715 \cite{Choudhury:2024xbk} as well as the 90\% credibility constraint on the tidal deformability from GW170817 \cite{TheLIGOScientific:2017qsa,Abbott:2018wiz}. We also require that the maximum mass configuration of the EoS supports at least $2\,M_{\odot}$ \cite{Demorest:2010bx,Antoniadis:2013pzd,Fonseca:2016tux,Cromartie:2019kug,Romani:2022jhd}. 
In general, stiff EoSs feature high maximum masses and large radii, compared to soft EoSs at lower masses and smaller radii. Therefore, stiff EoS are not strongly constrained by the $2.0\,M_{\odot}$ mass constraint or the radius estimate $R_{2\,M_{\odot}}\ge 11\,$km of a $2\,M_{\odot}$ star by NICER \cite{Miller:2021qha, Riley:2021pdl, Raaijmakers:2021uju}. However, the tidal deformability constraint obtained from GW170817 \cite{Abbott:2018wiz} requires compact solutions, i.e., a high value of the compactness, $C = M/R$. Soft EoSs tend to contain more compact neutron stars, as the mass decrease is smaller compared to the radius decrease when going from a stiff EoS to a soft one, which rules out particularly stiff EoSs. In our setup, the purely hadronic EoSs with $m^*/m < 0.65$ are ruled out completely, while only some parameter sets with $m^*/m = 0.65$ are consistent with the tidal deformability constraint.\\ 

\begin{figure}
		\centering				
		\includegraphics[width=0.4\textwidth]{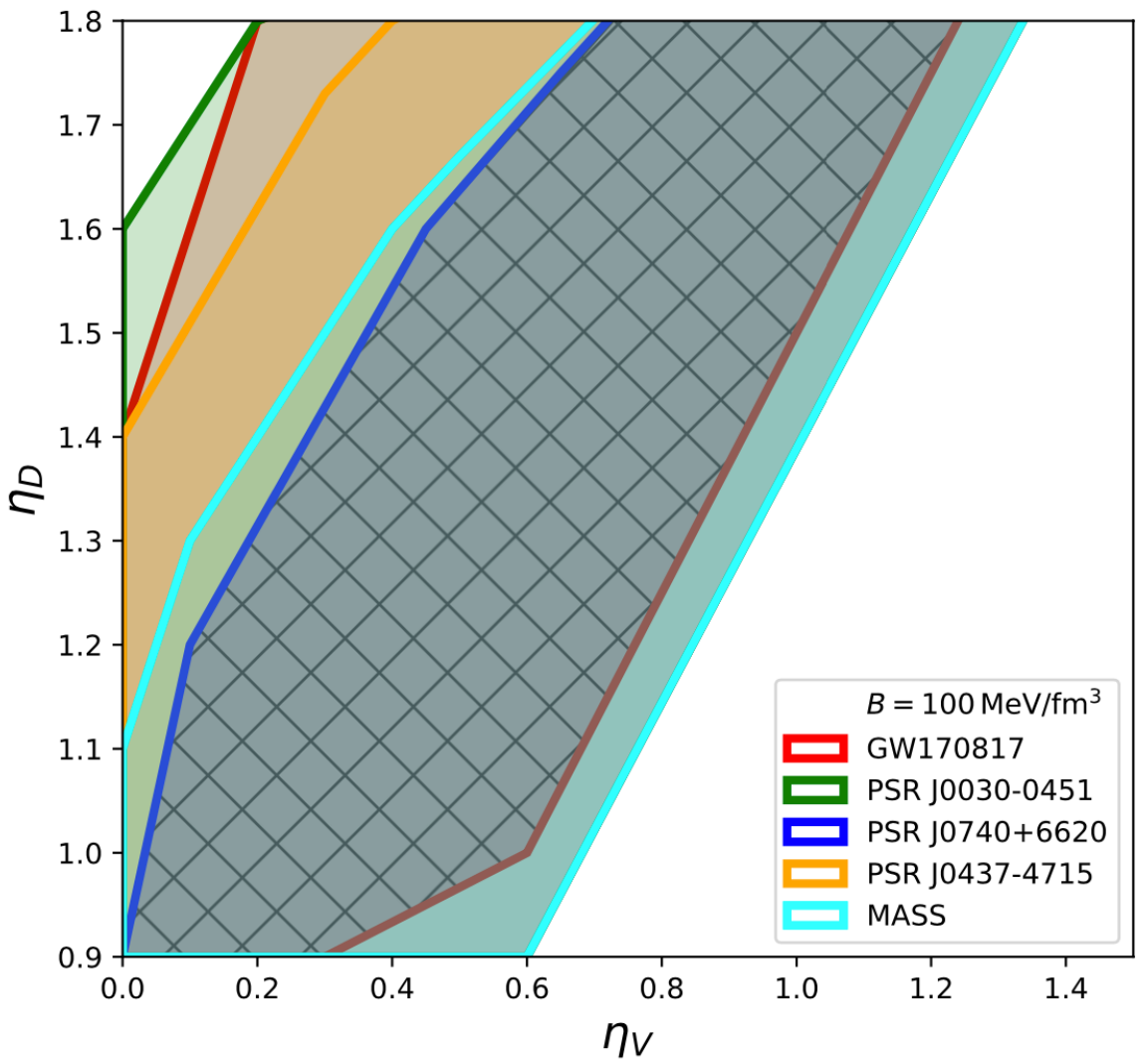}
		\caption{Values of the diquark coupling ($\eta_D$) and vector coupling ($\eta_V$) allowed by specific constraints for a hadronic EoS with $m^{*}/m=0.65$ and $B=100\,\mathrm{MeV/fm^3}$ that feature hybrid stars. The band where parameter sets exist that fulfill all constraints is cross-hatched. The specific regions to which $\eta_V$ and $\eta_D$ are constrained are indicated with colors that correspond to their respective observables. Red represents GW170817. Green, blue, and orange represent the NICER measurements. Finally, cyan represents the mass constraint. Note that all areas except for GW170817 share a border on the right side, beyond which no stable hybrid stars can be realized.}      
		\label{Fig:CAfm65}
\end{figure}

The way the observations outlined above constrain a given set of parameters is shown for the example of $m^{*}/m = 0.65$ and $B=100\,\mathrm{MeV/fm^3}$ in Fig.~\ref{Fig:CAfm65}. The shaded areas correspond to specific constraints. Similar to Fig.~\ref{Fig:CAfM}, not every parameter set containing the fixed parameters is guaranteed to be compatible with the constraint, but there is at least one parameter set for each point in the shaded area that is compatible.
A significant amount of purely hadronic EoSs with $m^{*}/m = 0.65$ are compatible with GW170817, even without a phase transition. Therefore, the GW170817 constraint (red) does not severely restrict the parameter space. The mass-radius estimate for PSR J0437-4715 (orange) has only slightly more constraining power than tidal deformability. The $2\,M_{\odot}$ mass constraint (cyan) is the most restrictive constraint and decreases the allowed range of $\eta_D$ for a given $\eta_V$ substantially. This can be further refined by including the radius measurement of the $2\,M_{\odot}$ PSR J0740+6620 (dark blue). Only the radius estimate for PSR J0030+0451 (green) limits the parameter space less than GW170817. All NICER constraints share a border on the right side of the plot.
This is not a limitation caused by astrophysical observations; instead, all parameter sets to the right of this border do not lead to stable hybrid stars and are therefore not of relevance to this work. 
Note that a plot similar to Fig.~\ref{Fig:CAfm65} can be made for a variety of parameter sets, but this one is well suited to outline the general trend: GW170817 requires softer hadronic EoSs, which are more difficult to combine with a quark phase. When already considering a soft hadronic EoS, the mass constraint becomes significantly more impactful than the tidal deformability constraint.\\

\begin{figure*}
		\centering				
		\includegraphics[width=0.8\textwidth]{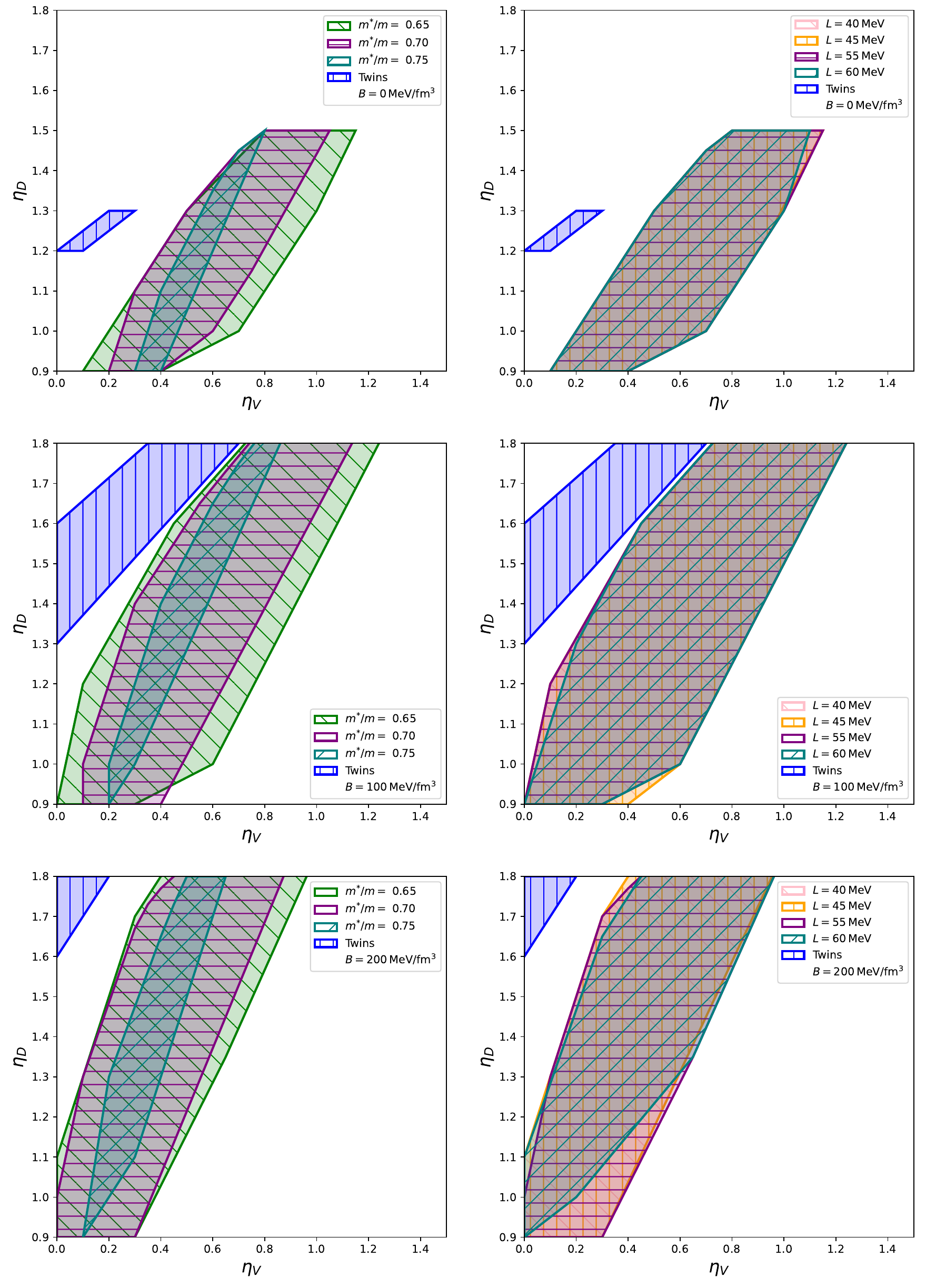}
		\caption{
        Left: The effective mass $m^{*}/m$ is kept constant, $n_0$, $J$, and $L$ are not fixed. The shaded areas correspond to the parameter sets that generate stars compatible with astrophysical constraints. Additionally, the blue-shaded area shows all parameter sets within the ranges of the used nuclear parameters that lead to twin stars.
        Right: The slope parameter $L$ of the hadronic phase is kept constant, $n_0$, $J$, and $m^{*}/m$ are not fixed. The results of the first row are for a bag constant $B=0\,\mathrm{MeV/fm^3}$, which is increased in each descending row by $100\,\mathrm{MeV/fm^3}$.
        }
		\label{Fig:CAfC}
\end{figure*}

\begin{table}[h]
    \centering
    \caption{The parameter sets (PS) used for Fig.~\ref{Fig:MRL}. The effective mass $m^*/m$, vector coupling $\eta_V$, and diquark coupling $\eta_D$ are dimensionless. Symmetry energy $J$ and slope parameter $L$ are in MeV. Saturation density $n_0$ and bag constant $B$ are in units of $\mathrm{fm^{-3}}$ and $\mathrm{MeV/fm^3}$, respectively. The transition density in the hadronic phase $n_{trans}$ is given in terms of $n_0$. The compatibility with GW170817 is marked with `yes/no`, as is the presence of twin stars. The last row indicates the quark phase at the hadron-quark transition.}
    \label{Tab:MR_Parametersets}
    \begin{tabular}{l c c c c c c}
        \toprule
        Parameter & PS1 & PS2 & PS3 & PS4 & PS5 \\
        \hline
        $m^*/m$  & 0.75 & 0.55 & 0.60 & 0.65 & 0.65 \\
        $L$ [MeV]  & 60 & 60 & 45 & 60 & 60 \\
        $J$ [MeV]  & 31 & 32 & 30 & 31 & 32 \\
        $n_0$ [$\mathrm{fm^{-3}}$] & 0.15 & 0.16 & 0.16 & 0.16 & 0.16 \\
        $\eta_V$  & 0.90 & 0.70 & 1.00 & 1.00 & 1.20 \\
        $\eta_D$  & 1.80 & 1.80 & 1.90 & 1.80 & 2.00 \\
        $B$ [$\mathrm{MeV/fm^3}$]  & 50 & 100 & 50 & 50 & 100 \\
        $n_{trans}$ [$n_0$] & 1.3 & 2.2 & 2.7 & 1.75 & 2.9 \\
        GW        & yes & no & no & yes & yes \\
        Twins     & no & yes & yes & no & no \\
        HM-QM transition     & 2SC & CFL & CFL & 2SC & CFL \\
        \hline
    \end{tabular}
\end{table}

Fig.~\ref{Fig:MRL} and Table~\ref{Tab:MR_Parametersets} suggest an incongruity between the tidal deformability constraint from GW170817 and twin stars. In Fig.~\ref{Fig:CAfC}, we investigate this thoroughly. The figure is structured identically to Fig.~\ref{Fig:CAfM}. The left side shows the $\eta_V-\eta_D$ parameter space in which constraints set above are met for at least one parameter set containing the effective mass corresponding to the color of the space. For the right side of the plot, $L$ is kept fixed. Only parameter sets that lead to stable hybrid stars are considered. 
The most noticeable influence on the allowed parameter range comes from the tidal deformability constraint, which rules out nearly all EoSs with $m^*/m \le 0.65$. For this reason, we only include EoSs with $m^*/m \ge 0.65$ in Fig.~\ref{Fig:CAfC}. However, stiffer EoSs are not completely ruled out. A phase transition to quark matter can, under certain circumstances, increase the compactness of the hybrid stars to such an extent that a previously ruled out hadronic EoS can regain compatibility with the tidal deformability constraint. In our framework, this effect is so small that we only found one combination of $\eta_V$ and $\eta_D$ that supports this phenomenon for a given bag constant. {This might be caused by a large step size for the coupling strength. Applying a smaller step size should allow us to find a small area rather than a single parameter combination currently identified. In our present study, the combination for $B=0\,\mathrm{MeV/fm^3}$ is $\eta_V=0.80$ and $\eta_D=1.50$ .
For these values, some parameter sets with $m^*/m = 0.55$ and $m^*/m = 0.60$ can fulfill all astrophysical constraints. However, stiff EoSs require a low transition density to be compatible with GW170817, in this case only slightly larger than $n_0$. Increasing the bag constant $B$ increases the transition density, necessitating corresponding adjustments to $\eta_V$ and $\eta_D$. At about $B\simeq100\,\mathrm{MeV/fm^3}$, the values for $\eta_V=1.00$ and $\eta_D=2.00$ are reached after which no transition at densities higher than $1\,n_0$ is possible for $m^*/m \le 0.60$ EoSs. Since small transition densities are required, all these parameter sets feature a realized 2SC phase. Conversely, this can be interpreted as a 2SC phase being required to ensure compatibility of an otherwise too stiff hadronic EoS with GW170817.\\

Furthermore, we find that by applying astrophysical constraints, the choice of the slope parameter $L$ loses significance. This is because the mass, radius, and tidal deformability are less sensitive to $L$ than to the effective mass, which gets restricted to such a large extent that all remaining parameter combinations are likely to be matchable no matter the choice of $L$. However, for very large $B$, this rule gets softened. We speculate that this is caused by the transition being moved to such large densities that even a slight stiffening in the hadronic EoS can shift the phase transition past the maximum mass configuration of the EoS.\\

Fig.~\ref{Fig:CAfC} also includes an area marked blue with parameter sets that lead to twin star solutions. The topic of twin stars only gains relevance in terms of their detectability. For this reason, we stipulate that observable twin star pairs needs to have a mass difference of $\Delta M \le 1\% \,M_{\text{twin}}$ and a radius difference of $\Delta R > 1\,$km, where $M_{\text{twin}}$ is the mass of the more massive twin. With this definition of twin stars, we are able to include mass-radius relations with a kink in addition to those with a separate second branch in our analysis.
To avoid visual confusion, the blue-shaded areas of Fig.~\ref{Fig:CAfC} are not further separated into different hadronic parameter sets. Since we already established that only one $\eta_V-\eta_D$ combination is possible for stiff EoSs, we only consider softer ones. In practical terms, this means that every point within the blue-shaded area has at least one parameter set with $m^*/m\ge0.65$ that leads to twin stars. This also means that points outside of this area cannot lead to twin stars as we defined them. Apart from that, no astrophysical constraints are applied to the twin star area. 
We find that under these conditions there is no overlap between the twin star area and the area constrained by astrophysical results. 
If we relax our definition of an observable twin star, for example by decreasing the required radius difference, this could be mitigated. 
However, twin stars are only relevant because they clearly point towards a phase transition. By reducing the required radius difference one would also reduce their differentiability from a purely hadronic EoS. One might also consider relaxing the constraint on the tidal deformability. For example, by using the original GW170817 tidal deformability and chirp mass values from Ref.~\cite{TheLIGOScientific:2017qsa} some parameter sets featuring twin stars, like PS3 in Fig.~\ref{Fig:MRL}, would be within the required bounds. However, when applying the updated 90\% credibility range from Ref.~\cite{Abbott:2018wiz}, this is not the case. In other words, the existence of twin stars in EoSs that reach two solar masses and more is strongly disfavored by GW170817, but not conclusively excluded.\\

\begin{figure*}[tbp]
		\centering				
		\includegraphics[width=0.9\textwidth]{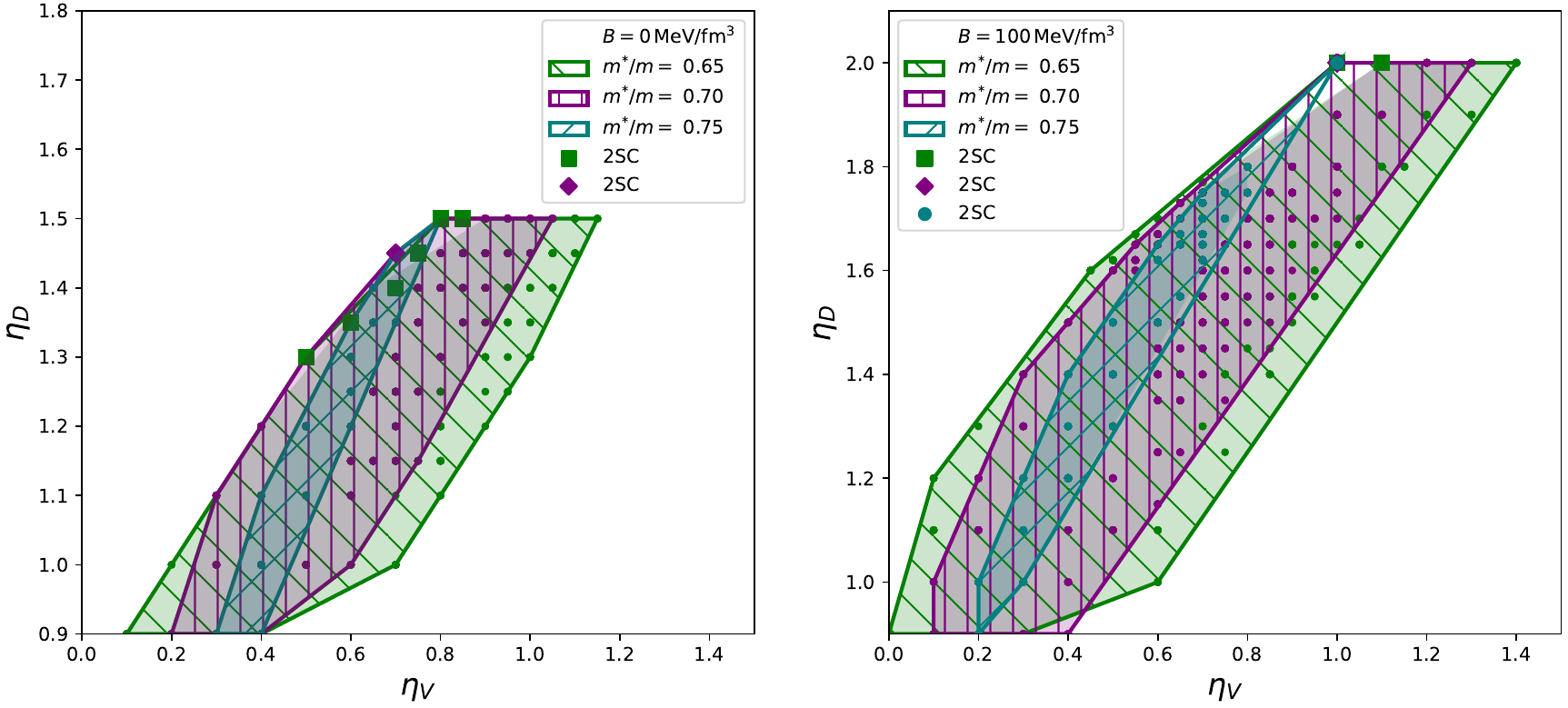}
		\caption{Constrained parameter space for varied effective masses with the quark phase at the point of the hadron-quark phase transition indicated. The shaded area represents where a transition to the CFL phase is possible for a parameter set containing the effective mass corresponding to the color of the region. The green square marker indicates where at least one hadronic parameter set with $m^*/m=0.65$ exhibits a phase transition to the 2SC phase. Likewise, the purple diamond marker represents a 2SC transition for $m^*/m=0.70$ and the teal circle for $m^*/m=0.75$. The small circles represent the investigated $\eta_V - \eta_D$ combinations within the constraint. The left side of the figure is for $B=0\,\mathrm{MeV/fm^3}$, the right side for $B=100\,\mathrm{MeV/fm^3}$.}
		\label{Fig:2Sc}
\end{figure*}

The parameter sets PS1-5 in Table~\ref{Tab:MR_Parametersets} might suggest that a large amount of EoSs contain 2SC quark matter, and for the unconstrained EoSs, this is true. 
However, most EoSs containing a 2SC phase are not compatible with astrophysical constraints, Instead, the majority of the constrained EoSs have a phase transition from the hadronic phase directly to the CFL phase. Only a small set of $\eta_V$ and $\eta_D$ combinations in the upper left edge of the allowed parameter space lead to both a 2SC and a CFL phase being realized within neutron stars. We show this in Fig.~\ref{Fig:2Sc}, where $\eta_V-\eta_D$ combinations that lead to a CFL phase are represented as small circles. The area in which they can be found is shaded, corresponding to the effective mass they need to be combined with, to be within the constrained area. The few points leading to a 2SC phase are indicated with large markers.
The absolute values of $\eta_V$ and $\eta_D$, and the number of possible combinations can change depending on the specific hadronic parameters, but their location in the allowed space remains the same. We find that the occurrence of a 2SC phase is highly dependent on the stiffness of the hadronic EoS, where stiffer hadronic EoSs exhibit a transition to a 2SC phase more often than softer ones. 
In contrast, by softening the hadronic EoS, the number of available combinations leading to a 2SC phase shrinks. For the largest effective mass, $m^*/m=0.75$, no stable 2SC phase is possible if the bag constant is $B=0\,\mathrm{MeV/fm^3}$.
This changes slightly by increasing $B$. If one disregards stiffness, the trend of increasing the bag constant is a reduction in available 2SC transitions. However, for a fixed stiffness of the hadronic EoS, increasing $B$ also moves the transition density to larger values. This means that some particularly soft EoSs can match to the quark phase just above saturation density if $B$ is increased.
Such is the case for $m^*/m=0.75$, which can have a 2SC transition when $B=100\,\mathrm{MeV/fm^3}$ but not for $B=0\,\mathrm{MeV/fm^3}$. 
This dependence on stiffness is the reason why including astrophysical constraints such as GW170817 excludes most EoSs with a 2SC phase. Should future evidence point towards even softer EoSs, a neutron star with a 2SC phase would become even more difficult to construct.\\

\section{Discussion and Outlook}\label{Sec:disc}

We performed a comprehensive parameter study for hybrid star EoSs composed of the FB RMF model combined with the renormalization group-consistent NJL (RG-NJL) model, including color superconductivity for the quark phase, making full use of their respective flexible frameworks. For the FB EoSs the investigated parameters are the symmetry energy $J$, the slope parameter $L$, and the effective mass $m^*/m$, all of which are fixed at saturation density $n_0$ and varied in ranges compatible with terrestrial experiments, as well as chiral effective field theory calculations. For the RG-NJL model, we varied the vector coupling strength $\eta_V$, diquark coupling strength $\eta_D$ as well as the bag constant $B$ within a large range. In this model, homogeneous phases with color superconductivity of three-flavor quark matter are calculated self-consistently. At high densities, matter is in the color-flavor-locked (CFL) phase, while at lower densities, two-flavor color superconductivity (2SC phase) is favored. 

The hadron-quark phase transition is modeled using a Maxwell transition and the 2SC-CFL transition is calculated self-consistently in the NJL model (neglecting 2SC-CFL mixed phases).\\

Depending on the used parameter set, we find hybrid stars with either a pure 2SC, a pure CFL, or both phases present in the quark matter core. However, all parameter sets have in common that the most compact (densest) stable star is composed of CFL matter in its center, even if a less dense star of the same parameter set might have a pure 2SC quark core. 
By applying astrophysical constraints in the form of mass, radius, and tidal deformability, the parameter ranges for possible matches are significantly reduced. For these constrained parameter sets, we found that the EoSs containing only a CFL phase are significantly more abundant than EoSs which can lead to 2SC quark matter.
This is because a phase transition to a 2SC phase in the EoS requires a stiff hadronic EoS,
however, stiff EoSs are disfavored by the tidal deformability constraint.\\

A major component of our analysis was the prospect of detecting the presence of quark matter in neutron stars. For this consideration, we focused on mass and radius as potential indicators. A nearly unambiguous indicator for 
 the presence of quark matter would be a so-called twin star, where two stars have the same mass but different radii.
A popular approach to investigate the possibility of quark matter in compact stars is the constant speed of sound (CSS) approach \cite{Alford:2013aca}. It is significantly less microphysically motivated than the framework we employ here, but it can be very useful for generalized investigations, especially when looking for twin stars. 
Using this approach Ref.~\cite{Li:2024lmd} find hybrid stars with maximum masses of more than $2\,M_\odot$ and within NICER as well as tidal deformability limits. This requires a high speed of sound, $c_s^2\ge2/3$ and often features a phase transition at small densities impeding detectable twin stars.
Ref.~\cite{Christian:2023hez} found that even under idealized circumstances ($c_s^2=1$) only a small set of phase transitions can lead to twin stars. 
Those transitions usually require a hadronic EoS of moderate stiffness and need to take place at high densities (see also Ref.~\cite{Christian:2021uhd}). 
The CSS quark model can also be used for sequential phase transitions, which should more closely resemble the framework employed here. For example, by setting the speed of sound to $c_s^2=0.7$ and $c_s^2=1$ for the 2SC and CFL phase respectively Ref.~\cite{Li:2023zty} finds a configuration where the 2SC-CFL transition induces another unstable region, resulting in potentially three stars with the same mass but different radii, so-called triplets. However,
as was shown in Ref.~\cite{Gholami:2024diy}, the speed of sound in the RG-consistent NJL model is not a constant quantity but can be non-monotonic at the 2SC-CFL transition. Furthermore, it only reaches values of $c_s^2\approx 0.5-0.6$ at densities in the cores of neutron stars. This reduced speed of sound compared to the CSS approach hinders the construction of twin or triplet stars. As a result, it is not possible to construct twin or triplet stars within the parameter ranges allowed by astrophysical constraints. While twin stars can be constructed outside these constraints by using a combination of smaller values for the vector coupling and larger values of the diquark coupling, triplet stars cannot be constructed at all, despite the 2SC-CFL transition being present in multiple parameter sets.\\

In summary, we find large parameter ranges for both hadronic and quark parameters that are compatible with all astrophysical constraints. This includes parameter sets with various transition densities, as well as parameter sets featuring both, 2SC and CFL quark matter. Despite this we do not find any clear mass-radius indicators that could help differentiate between hybrid and purely hadronic stars \cite{Alford:2004pf}.
Furthermore we find that the occurrence of a 2SC phase correlates strongly with a stiff hadronic EoS. Consequently, any future observational evidence indicating a softer EoS would significantly restrict the possibility of sequential phase transitions.\\

Our work can be extended in various ways. In the present study, we concern ourselves with the allowed parameter ranges for a potential phase transition, we do not consider whether or not a phase transition is likely. For this reason, a full Bayesian analysis might elevate our work and provide a likelihood function for the presence of quark matter in neutron stars \cite{Pfaff:2021kse, Annala:2023cwx, Albino:2024ymc, Kurkela:2024xfh}. Furthermore, we suggest looking more strongly at alternative ways of detecting the presence of a phase transition such as cooling behavior \cite{deCarvalho:2015lpa, Lyra:2022qmg, Mendes:2022gbq, Mendes:2024hbn}, oscillation modes \cite{Ranea-Sandoval:2018bgu, Pradhan:2023zmg, Rather:2024nry} or thermal twins \cite{Hempel:2015vlg, Carlomagno:2023nrc}.\\

    \begin{acknowledgments}
The authors thank Michael Buballa, Alexander Haber, and Jürgen Schaffner-Bielich for helpful discussions. JEC was funded by the European Research Council (ERC) Advanced Grant INSPIRATION under the European Union's Horizon 2020 research and innovation programme (Grant agreement No. 101053985). IAR acknowledges support from the Alexander von Humboldt Foundation. HG acknowledges support from the Deutsche Forschungsgemeinschaft (DFG, German Research Foundation) 
through the CRC-TR211 'Strong-interaction matter under extreme conditions' project number 315477589 – TRR 211. MH is supported by the GSI F\&E. 
    \end{acknowledgments}
    
    \bibliographystyle{apsrev4-1}
	\bibliography{neue_bib_22}

\end{document}